\documentclass[reqno,12pt]{amsart}

\usepackage{enumerate}
\usepackage[margin=1in]{geometry}
\usepackage{enumitem}
\usepackage{ifpdf}
\usepackage{amsmath}
\usepackage{amsfonts}
\usepackage{amssymb}
\usepackage{amsaddr}
\usepackage{amsthm}
\usepackage{amsrefs}
\usepackage{mathrsfs}
\usepackage{cite}
\usepackage[hyperfootnotes=false]{hyperref}
\usepackage[dotinlabels]{titletoc}
\usepackage{nicefrac}
\usepackage{float}
\usepackage[multiple]{footmisc}
\usepackage{graphicx}

\newcommand{\innprod}[2]{\langle #1 | #2 \rangle}

\newcommand{\norm}[1]{\left\lVert {#1} \right\rVert}
\newcommand{\ket}[1]{\ensuremath{\left|#1\right\rangle}}

\author{Asif Shakeel$^\dag$,  David A. Meyer$^\ddag$ and Peter J. Love$^\dag$}
\address{$^\dag$Department of Physics, Haverford College, Haverford, PA 19041, USA}
\address{$^\ddag$Department of Mathematics, University of California/San Diego, La Jolla, CA 92093-0112, USA}
\email{ashakeel@haverford.edu, dmeyer@math.ucsd.edu, plove@haverford.edu}
\date{\today}

\ifpdf
\fi

\title[History dependent QRW as QLGA] {History Dependent Quantum Random Walks as Quantum Lattice Gas Automata}

\begin{document}

\begin{abstract}
Quantum Random Walks (QRW) were first defined as one-particle sectors of Quantum Lattice Gas Automata (QLGA).  Recently, they  have been generalized to include history dependence, either on previous coin (internal, i.e., spin or velocity) states or on previous position states. These models have the goal of studying the transition to classicality, or more generally, changes in the performance of quantum walks in algorithmic applications. We show that several history dependent QRW can be identified as one-particle sectors of QLGA. This provides a unifying conceptual framework for these models in which the extra degrees of freedom required to store the history information arise naturally as geometrical degrees of freedom on the lattice. 
\end{abstract}

\maketitle

\section{Introduction} \label{section:intro}

Classical random walks~\cite{bib:pearson} provide a discrete model for the heat/diffusion equation, in which the variance of position after $T$ steps scales as $T^{1/2}$~\cite{bib:raleigh1880,bib:raleigh1905,bib:bachelier,bib:polya1,bib:polya2}.  Beyond their physical significance, random walks also have a host of algorithmic applications, ranging from Monte Carlo methods~\cite{bib:metropolisulam} to probabilistic \textsc{Satisfiability} solvers~\cite{bib:papadimitriou,bib:schoning}.

Similarly, {\sl quantum} random walks (QRW) provide discrete models for the Dirac equation~\cite{bib:meyer2} in which the variance in position after $T$ steps scales as $T$~\cite{bib:abnvw}.\footnote{Note that the Dirac equation is a $\Delta x/\Delta t$ constant continuum limit of the QRW in 1 dimension.  Taking a $(\Delta x)^2/\Delta t$ constant continuum limit gives the Schr\"odinger equation~\cite{bib:boghosian1998} which has position variance scaling as it does in the diffusion equation (which is the same continuum limit of the classical random walk).} Furthermore, Grover's quantum search algorithm~\cite{bib:grover} can be understood as a QRW in the presence of a potential, and QRW search algorithms have been found for hypercubes~\cite{bib:skw} and for cubic lattices in various dimensions~\cite{bib:akrcmqwf}.  Continuous time analogs of random walks are particularly natural in the quantum setting, where Hamiltonian evolution can be defined~\cite{bib:farhigutmann}.  These give algorithms for spatial search~\cite{bib:childsgoldstone1,bib:childsgoldstone2} and NAND tree evaluation~\cite{bib:farhigoldstonegutmann}, and provide an example of a provable exponential speedup for the problem of traversing a pair of binary trees connected by a random cycle~\cite{bib:childesclevedeotto}.  

Classical random walks have been generalized to retain a history of previous positions.  The self-avoiding random walk was introduced to model configurations of polymers~\cite{bib:orr,bib:flory}.  Even before the model was formalized, it was recognized that the self-avoidance condition would lead to increased scaling of the linear size of the conformation with the length of the polymer and hence increased viscosity~\cite{bib:kuhn,bib:flory,bib:floryfox}.  Nothing has been proved about this scaling (above dimension 1), however, but a closely related model, the loop-erased random walk, has been defined and is more amenable to analysis~\cite{bib:lawler}.  Consequently the latter has been incorporated into improved algorithms for uniformly sampling spanning trees in a graph~\cite{bib:pemantle} and more generally into the Propp-Wilson algorithm for exact sampling from the equilibrium distribution of a Markov chain~\cite{bib:wilson,bib:proppwilson}.  

Given what is known about classical and quantum random walks, and about classical random walks with memory, it is very natural to investigate quantum counterparts of the latter. Several groups have done so, proposing a variety of models for history dependent QRW~\cite{bca:qwdmc,bca:qctrw,mg:odqwwm,rbg:qwwmrccf,pbhmpbk:nrnrqw,crt:saqw}, including one (in continuous time~\cite{bib:Rosmanis}) explicitly motivated by the algorithmic problem of finding a connecting path in the glued binary trees problem mentioned above.  The standard QRW in one dimension was first defined as the single particle sector of a quantum lattice gas automaton (QLGA)~\cite{bib:meyer2}.  Recently, more general multiparticle quantum lattice gas models have been studied as special cases of quantum cellular automata (QCA)~\cite{bib:slwqcaqlga}.  In this paper we combine these observations to show that each of the history dependent QRW defined to date is, in fact, a QLGA and thus, {\it a fortiori}, a QCA.  Beyond providing a unifying framework within which to construct and compare these models, QCA, with their physical interpretation as discretized quantum field theories, suggest possible physical meanings, and even instantiations, of history dependent QRW.

The paper is organized as follows. In Section~\ref{sec:qrw} we introduce discrete quantum random walks, and describe the extensions of these models to include particle and site history dependence. In Section~\ref{sec:qlga} we introduce the quantum lattice-gas automata before formulating QRW with history dependence as QLGA in Section~\ref{sec:qrwqlga}. Finally, we consider non-reversing and non-repeating walks as two-dimensional QLGA in Section~\ref{sec:nrnr}.  We close the paper with some discussion and directions for future work. 

\section{QRW} \label{sec:qrw}

We first describe a QRW and then include the  history dependence as in~\cite{bca:qwdmc,bca:qctrw,mg:odqwwm,rbg:qwwmrccf,pbhmpbk:nrnrqw,crt:saqw,bib:Rosmanis}. A complete basis for the Hilbert space of a discrete time QRW is labelled by the position and a velocity that corresponds to the direction in which the particle has a tendency to move~\cite{bib:schrodinger1,bib:schrodinger2}.  In one dimension the position space is $\ell^2(\mathbb{Z})$ (or some periodic quotient thereof) with computational basis elements $\ket{x}$,  $x \in \mathbb{Z}$, and the inner product norm.  The inner product of elements $\alpha = \sum_x \alpha_x \ket{x}$ and $\beta = \sum_x \beta_x \ket{x}$ is
\begin{equation*}
\innprod{ \alpha}{\beta} = \sum_x \bar\alpha_x {\beta_x}.
\end{equation*}
The velocity space for the simplest models in one dimension has computational basis $\ket{p}$, $p\in\{+1,-1\}$, and hence is a two-dimensional space (the two dimensions corresponding to left and right directions on a one dimensional lattice) $\mathbb{C}^2$, with the inner product and the norm finite versions of the above.  The QRW Hilbert space $\mathcal{H}$ is
\begin{equation*}
\mathcal{H} = \ell^2(\mathbb{Z}) \otimes\mathbb{C}^2.
\end{equation*}
By construction, $\mathcal{H}$ has an inner product, and its accompanying norm, induced from those on the component spaces (product of the respective inner products on basis elements, extended by linearity to $\mathcal{H}$). The state of a QRW is an element $\psi \in \mathcal{H}$ of unit norm $\norm{\psi} = 1$.

A QRW evolves through two consecutive unitary actions on its state:
 \begin{enumerate}[label=(\roman{*})] 
\item \label{prop} Advection
\begin{equation*}
A : \ket{x}   \ket{p} \mapsto \ket{x+p}   \ket{p}.
\end{equation*}
\item \label{scat}  Scattering
\begin{equation*}
I\otimes S : \ket{x} \ket{p} \mapsto \ket{x} S(\ket{p}),
\end{equation*}
where $S$ is a unitary map on $\mathbb{C}^2$.  The resulting change in the two-dimensional velocity tensor factor is analogous to a coin flip, so this tensor factor is often described as a ``coin''.  Since we are emphasizing physical interpretations in this paper, we will refer to it as velocity~\cite{bib:schrodinger1,bib:schrodinger2}. 
\end{enumerate}   
The QRW transition $U$ is
\begin{equation*}
U = A (I\otimes S).
\end{equation*}
In~\cite{bib:meyer2}, the symmetric scattering matrix,
\begin{equation} \label{velocityun}
S=\begin{pmatrix}
\cos\theta& i  \sin\theta \\
i \sin\theta &\cos\theta \\
\end{pmatrix},
\end{equation}
is used to derive the Dirac equation as a continuum limit of the QRW.
History dependent QRW~\cite{bca:qwdmc,bca:qctrw,mg:odqwwm,rbg:qwwmrccf,pbhmpbk:nrnrqw,crt:saqw,bib:Rosmanis} build upon this basic model in two ways, which we classify as {\it particle\/} and {\it site\/} history dependence.

\subsection{QRW with particle history dependence} \label{subsec:qrwhistc}
Particle history refers to models in which a particle, as it scatters from a site, updates a   finite history stored in extra internal degrees of freedom,  without any effect on the site.  The sites can be described as being {\it inert}. The history that the particle carries then influences its future scattering. In the standard QRW this is a one-step history preserving only the memory of the velocity from the previous scattering. In general, this history can be a ``filtered'' version of the previous scatterings, preserved in some finite dimensional Hilbert space that is part of the particle's internal state. The general Hilbert space in this case would be
\begin{equation*}
\mathcal{H} = \ell^2(\mathbb{Z}) \otimes  V_p, 
\end{equation*}
for some finite-dimensional {\it particle history\/} Hilbert space $V_p$. 
Scattering at a site then affects the tensor factor $V_p$ and the advection is an operation on the tensor factor $\ell^2(\mathbb{Z})$ controlled by the state of $V_p$.

We can fit most of the history dependent QRW that have been defined previously into this framework. \begin{enumerate}[label=(\alph{*})] 
\item \label{scmem} Several previous models maintain a history comprising a length $N$ ``tail'' of  previous velocities~\cite{bca:qwdmc,bca:qctrw,rbg:qwwmrccf}, recorded in $N$ qubits appended to the QRW position. These are updated at each time step.  The current velocity is chosen in a cyclic manner from the multiple ($N$) velocities.  At time step $k$, where $1\leq k \leq N$, that velocity  is  scattered by a scattering operator $S_k$, and the resulting velocity is used in the advection step.
\item \label{snmem} Another approach is to construct a model which maintains a history comprising a length $N$ ``tail'' of sites visited prior to the current site~\cite{mg:odqwwm}.  The step operation, i.e., the next right or left site, is determined in a two stages:  by flipping a coin (quantum) and then scattering the current velocity with a symmetric matrix that either ``reflects'' (reverses the previous velocity) or ``transmits'' (maintains the previous velocity), according to the coin outcome.

\end{enumerate}   

It is clear that case~\ref{snmem} can be mapped to case~\ref{scmem} as far as the history of previous sites is concerned, since a history of sites can be rephrased as a history of previous velocity values. We therefore only consider velocity memory. In the case~\ref{scmem} models~\cite{bca:qwdmc,bca:qctrw,rbg:qwwmrccf}, in order to store the history, the single velocity qubit of a QRW is replaced with multiple velocity qubits,  $\bigotimes^N \mathbb{C}^2$.  The Hilbert space for a QRW with particle history is then
\begin{equation*}
\mathcal{H} = \ell^2(\mathbb{Z}) \otimes  \bigotimes^N \mathbb{C}^2 =  \ell^2(\mathbb{Z}) \otimes  V_p.
\end{equation*}
Computational basis elements of the Hilbert space $\mathcal{H}$ can be taken as
\begin{equation*}
 \ket{x}  \ket{p_1\ldots p_{N}},
\end{equation*}
where $\ket{x}$ is the position,   and $\ket{p_k} \in\{\ket{+1},\ket{-1}\}$, $1\leq k \leq N$, hold the history of velocities.

The memory of the velocities is updated each time a velocity is used to determine the next site. The QRW transition in Brun, Carteret and Ambainis' model~\cite{bca:qwdmc} and in Rohde, Brennen and Gilchrist's model~\cite{rbg:qwwmrccf} (both described by case~\ref{scmem} above) scatters the  velocity that happened $N$ steps ago. Thus it is convenient to split the scattering into two distinct stages as in~\cite{rbg:qwwmrccf}, the first of which is  the selection of the velocity to scatter or the {\it memory\/} operation and the other is the {\it ricochet\/} operation that refers to the interaction between the particle and the site. Since the site is inert, this is also described as the {\it self-interaction\/} of a particle.  The operations in use are:
 \begin{enumerate}[label=(\roman{*})] 
\item \label{memo2}  Memory  
\begin{equation*}
M : \ket{p_1 p_2 \ldots p_{N}} \mapsto  \ket{p_N p_1 p_2 \ldots p_{N-1}}.
\end{equation*}
\item \label{scat2}  Ricochet  
\begin{equation*}
R :  \ket{p_1 p_2 \ldots p_{N}} \mapsto   \ket{p_1 p_2 \ldots p_{N-1}}R_k(\ket{p_{N}}),
\end{equation*}
where $R_k$, $1\leq k \leq N$, is the symmetric scattering matrix, parameterized by $\theta_k$ for the $k$-th velocity selection  (all $R_k$ are identical in~\cite{rbg:qwwmrccf}),
\begin{equation*} 
 R_k=\begin{pmatrix}
 \cos\theta_k&i \sin\theta_k\\
i \sin\theta_k&  \cos\theta_k\\
\end{pmatrix}.
\end{equation*}
\item \label{prop2} Advection, $A$, which is   the shift of the current position by the appropriate velocity value,
\begin{equation*}
A : \ket{x}    \ket{p_1 p_2 \ldots p_{N}}  \mapsto \ket{x + p_{N}}  \ket{p_1 p_2  \ldots p_{N}}.
\end{equation*}
\end{enumerate}   

The QRW transition in~\cite{rbg:qwwmrccf}, let us call it $U$, is given  as
\begin{equation*}
U =  (I \otimes M) A  (I \otimes R).
\end{equation*}
And it is clear that a simple redefinition of $A$ above to 
\begin{equation*}
\tilde A : \ket{x}     \ket{p_1 p_2 \ldots p_{N}}  \mapsto \ket{x + p_1}  \ket{p_1 p_2  \ldots p_{N}},
\end{equation*}
leads to
\begin{equation*}
U =   \tilde A  (I \otimes M R).
\end{equation*}
The scattering operator would now be
\begin{equation*}
S = M R. 
\end{equation*}
So we get
\begin{equation*}
U =   \tilde A  (I \otimes S).
\end{equation*}

In Mc~Gettrick's model~\cite{mg:odqwwm} (described in case~\ref{snmem} above), the QRW transition  is derived by a different method which is not achievable by simple cyclic rotation of coins, since it keeps a history of previous positions as it moves to the next one.  Once phrased in terms of the current position and previous velocities, the Hilbert space on which this QRW is defined is
\begin{equation*}
\mathcal{H} = \ell^2(\mathbb{Z}) \otimes  \mathbb{C}^2   \otimes  \bigotimes^N \mathbb{C}^2 =  \ell^2(\mathbb{Z}) \otimes V_p.
\end{equation*}
Written  in this form, we can  designate the first $\mathbb{C}^2$ tensor factor  of $V_p$ as the ``control" variable needed in determining the mode of scattering and the remaining $\mathbb{C}^2$ factors are the record of previous velocities.  The computational basis elements of the Hilbert space $\mathcal{H}$ are
\begin{equation*}
 \ket{x}    \ket{c} \ket{p_1\ldots p_{N}},
\end{equation*}
where $\ket{x}$ is the position,  $\ket{c}\in\{\ket{0},\ket{1}\}$ is the control state,  and $\ket{p_k} \in\{\ket{+1},\ket{-1}\}$, $1\leq k \leq N$, hold the history of velocities.

The model studied in~\cite{mg:odqwwm} uses a history with one previous velocity, $N=1$, or equivalently one previous  position (two positions in total). We give a simple generalization of  the scattering scheme used there to  an arbitrary number of velocities $N$. There is a memory operation as in the particle history dependence just considered, performing a circular shift of the velocity history. Then there is a two-stage operation.  Its first stage is the {\it ricochet-control\/} whose outcome  determines the next ricochet mode.  This ricochet-control is a simple symmetric scattering on the $ \ket{c}$ factor (control variable).  The next stage is the  {\it ricochet\/} operation which  is a controlled action by the control variable on the velocity part of the space to determine the velocity for the next move.  The operations in use for this scheme are:
 \begin{enumerate}[label=(\roman{*})] 
\item \label{memo3}  Memory
\begin{equation*}
M :     \ket{c} \ket{p_1 p_2 \ldots p_{N-1} p_N}  \mapsto \ket{c} \ket{p_N p_1 \ldots p_{N-2} p_{N-1}}.
\end{equation*}
\item \label{scon3}  Ricochet-Control  
\begin{equation*}
C :   \ket{c}  \ket{p_1 p_2 \ldots p_N} \mapsto U_s(\ket{c})\ket{p_1 p_{2} \ldots p_{N}},
\end{equation*}
where $U_s$ is a symmetric scattering matrix.
\item \label{sbr3}  Ricochet  
\begin{equation*}
R :   \ket{c} \ket{p_1 p_{2} \ldots p_{N}} \mapsto \ket{c}U_c(\ket{p_1 p_2})\ket{p_2 \ldots p_{N}},
\end{equation*}
where $U_c$ is a controlled symmetric scattering matrix  controlled by $\ket{c}$. In~\cite{mg:odqwwm}, both for $c=0$, and $c=1$, $U_c$ is fixed to be either identity (transmitting) or $X$ (reflecting). The particular case of $N=1$  for which the rules are given in~\cite{mg:odqwwm},  has the term $\ket{p_1 p_2}$ in the above  expression  as $\ket{p_1}$. 
\item \label{prop3} Advection
\begin{equation*}
A : \ket{x} \ket{p_1 p_2 \ldots p_{N}} \mapsto \ket{x + p_1}  \ket{p_1 p_2 \ldots p_{N}}.
\end{equation*}
\end{enumerate}   
The transition $U$ is
\begin{equation*}
U =   A   (I \otimes R  C M).
\end{equation*}
Writing the  scattering as  $S = R  C M$,  we again get the usual form for the transition,
\begin{equation*}
U =    A  (I \otimes S).
\end{equation*}

At the end of his paper, Rosmanis~\cite{bib:Rosmanis} suggests that his continuous time quantum ``snake walk'' should have a discrete time version.  Were we to construct it in the one dimensional case, it would have some similarities with Mc~Gettrick's model~\cite{mg:odqwwm}:  there is a string of adjacent locations comprising the head to tail of the ``snake'', which would be treated as a particle history attached to the head position, and the ``snake'' can move its head, or its tail---this decision would be implemented with a four dimensional unitary transformation like the ricochet operation above, after which there would be the corresponding forward or backward, left or right, advection (``slither''!).

\subsection{QRW with site history dependence} \label{subsec:qrwhistn}
In contrast to particle history, Camilleri,\break Rohde and Twamley consider a different kind of history dependence in~\cite{crt:saqw}, the history of site visits.  This history is the result of the particle changing the state of a site when it scatters from it.  In this sense, the sites are {\it active}. The state of each site then affects how the particle scatters at the site in the future. 
 
Just as in the particle history, one might like to retain the history of each site by an identical  finite dimensional Hilbert space $V_s$ per site. Assuming that there are $N$ sites, i.e.,  the lattice has size $N$, this requires appending a space  $\bigotimes^N V_s$ to keep track of the  site history of every site.  The QRW Hilbert space is then
\begin{equation}\label{eq:fqsq}
\mathcal{H} = \ell^2(\mathbb{Z}_N) \otimes V_p \otimes  \bigotimes^N V_s,
\end{equation}
where $V_p$ is the particle history Hilbert space as before.

In~\cite{crt:saqw}, there is a one qubit memory of visits for each site, indicating if the site has or not been visited in the past.  Then the QRW Hilbert space becomes
\begin{equation*}
\mathcal{H} = \mathbb{C}^N\otimes\mathbb{C}^2\otimes  \bigotimes^N  \mathbb{C}^2.
\end{equation*}
In this case $V_p = \mathbb{C}^2$ and $V_s = \mathbb{C}^2$. 

A position is given by an element of $\mathbb{C}^N$, with a computational basis composed of  elements $\ket{x}$, $x \in \mathbb{Z}_N$.  The velocity space $V_p$ has the computational basis $\{\ket{+1},\ket{-1}\}$.  The record of a visit to the  site $x$ is in the corresponding  $x$-th  tensor factor (qubit) of  $\bigotimes^N  \mathbb{C}^2$,  with the basis of  $\mathbb{C}^2$ in that factor taken as $\ket{0}$,  denoting the ``not visited" state and $\ket{1}$,  denoting the ``have visited" state. 
A basis element of the state of all the memory qubits is
\begin{equation*}\ket{m_1 \ldots m_N} \in \bigotimes^N  \mathbb{C}^2,
\end{equation*}
where $m_x$  denotes the memory qubit  corresponding to site $x \in \mathbb{Z}_N$. We, therefore,  write a computational basis elements of the  Hilbert space $\mathcal{H}$ as
\begin{equation*}
 \ket{x} \ket{p}  \ket{m_1  \ldots m_N}.
\end{equation*}
The memory qubits are a record of the site history, hence are part of a different  Hilbert space that cannot be grouped together with the particle history.  The memory for a site  is updated each time that site  is visited,  and the scattering of a particle at that site is controlled by the state of the corresponding memory qubit. Thus,   unlike the case of particle history,  scattering cannot be said to act on the ``velocity space'' independently of the other parts.  The operations included in a  transition are:
\begin{enumerate}[label=(\roman{*})] 
\item \label{memo3}  Memory (implicitly controlled by $\ket{x}$),  
\begin{equation*}
M :   \ket{x} \ket{p}  \ket{m_1  \ldots m_N} \mapsto \ket{x}   \ket{p}  \ket{m_1 \ldots m_{x-1}}U_M(\ket{m_x}) \ket{m_{x+1}\ldots m_N},
\end{equation*}
where $U_M$ is the symmetric scattering matrix, parameterized by $\theta_M$ (the memory strength),
\begin{equation} \label{memun}
 U_M=\begin{pmatrix}
 \cos\theta_M&i \sin\theta_M\\
i \sin\theta_M&  \cos\theta_M\\
\end{pmatrix}.
\end{equation}
\item \label{scat2}  Ricochet, a controlled operation on the $\ket{p}$ factor,   controlled by the qubit $\ket{m_x}$ (implicitly controlled by $\ket{x}$ as well),  
\begin{equation*}
R :   \ket{x}  \ket{p}  \ket{m_1 \ldots m_x \ldots m_N}  \mapsto \ket{x}   U_{m_x}(\ket{p})  \ket{m_1 \ldots m_x \ldots m_N},
\end{equation*}
where $U_{m_x}$, for  $m_x = 0$,  is the ``balanced'' walk,
\begin{equation} \label{fortun}
 U_{0}=\frac{1}{\sqrt{2}}\begin{pmatrix}
 1&i \\
i &  1\\
\end{pmatrix},
\end{equation}
and  for  $m_x = 1$, it is the symmetric scattering matrix, parameterized by $\theta_b$ (the ``back action''),
\begin{equation} \label{bactun}
 U_{1}=\begin{pmatrix}
 \cos\theta_b&i \sin\theta_b\\
i \sin\theta_b&  \cos\theta_b\\
\end{pmatrix}.
\end{equation} 
\item \label{prop3} Advection 
\begin{equation*}
A : \ket{x}   \ket{p}  \ket{m_1 \ldots m_N}  \mapsto \ket{x+p}   \ket{p} \ket{m_1 \ldots m_N}.
\end{equation*}
\end{enumerate}    
The QRW transition in~\cite{crt:saqw} is given as composition $U$,
\begin{equation*}
U =  A   R M.
\end{equation*}
Since $R$ and $M$ are non-identity on every factor, we use the term ``generalized scattering'' denoted $\hat S$, to be $\hat S = R M$. Then the transition becomes
\begin{equation*}
U =  A   \hat S.
\end{equation*}

Notice that $U_0$ is $U_1(\theta_b = \pi/4)$; while it is natural for these scattering matrices to be symmetric, since that encodes parity invariance (left-right symmetry), the obvious generalization is to simply use two different values of $\theta$ in the $S$ matrix of Eq.~\eqref{velocityun} for the two cases. 
 Moreover, since $\theta$ goes to the (effective) mass in the continuum limit~\cite{bib:meyer2}, these choices have a physical meaning.  Notice also that the site history comprises a ``parallel'' memory in the sense that since the particle position is a superposition of sites at each timestep, the entire ``checker-board'' of site history is now entangled with the position and particle history.  A natural interpretation of these {\it active\/} sites is as a lattice of immobile particles, a discrete background quantum field~\cite{bib:jlp}, that interacts with the hopping particle of the QRW.  We note, in passing, that this is essentially a quantum mechanical version of the classical Lorentz (or wind-tree) lattice gas model introduced in~\cite{bib:ruijgrokcohen}.  Furthermore, since there is no necessary parity invariance in the interaction of the moving particle with an immobile particle $U_M$ may be generalized to an arbitrary unitary matrix, not necessarily symmetric.

This {\it active\/} site interpretation suggests a multi-particle picture in which one is concerned with the presence or absence of particles at all the sites simultaneously, viewing the  checker-board as the state of a set of static particles, rather than keeping track of the position and velocity of a single particle.  The evolution is local in the sense that only when a hopping particle arrives at a site can it bounce off the static particle and go to one of the neighbors, affecting the site and itself.  Notice that in the QRW with site history example just given, the Hilbert space dimension grows exponentially, as $N 2^{N+1}$. This is not the recipe for a useful Hilbert space with a nice topology; rather, one might want to control the Hilbert space dimension.  These ideas are incorporated in a multiparticle generalization of the types of QRW we have just discussed:  Quantum Lattice Gas Automata (QLGA)~\cite{bib:meyer2,bib:slwqcaqlga}, which we describe in the next section.

\section{QLGA}\label{sec:qlga}
A QLGA is a  model of particles on a lattice propagating and scattering  by interactions when they arrive at  lattice sites (or cells)~\cite{bib:meyer2,bib:slwqcaqlga}. Each site/cell has a set of neighboring cells  and the particles are exchanged with those neighbors. 

We consider the QLGA model as formulated in~\cite{bib:slwqcaqlga}. Take the lattice of cells to be $\mathbb{Z}$.  Each cell can be occupied by multiple particles, and each particle has a state which is a vector in a {\it subcell\/} Hilbert  space.  Let us say that the internal states of particle $j$, $1 \leq j \leq d$, are  elements of the subcell Hilbert space $W_j$, so the cell Hilbert space is $W = \bigotimes_{j=1}^d W_j$.  In the propagation stage of the evolution, a particle with internal state $j$ (or equivalently, that occupies subcell state $j$) hops to  the corresponding subcell $j$ of a designated neighboring cell, $e_j \in \mathbb{Z}$ away.  The collection of such neighbors is specified by the {\it neighborhood\/} $\mathcal{E} = \{e_1,e_2,\ldots,e_d\}\subset\mathbb{Z}$, of cardinality $|\mathcal{E}| = d$.  

Let $\mathcal{B}$ be a basis  of the cell Hilbert space $W$ expressed in terms of some orthonormal basis $\mathcal{B}_j $ of $W_j$, 
\begin{equation*}
\mathcal{B} = \{ \ket{b} = \bigotimes_{j=1}^{d} \ket{b_j} :   \ket{b_j} \in  \mathcal{B}_j\}.
\end{equation*}
 If we index the Hilbert space of cell $x$ as $W^x$, and the corresponding basis $\mathcal{B}^x$, then we can carry this index into the subcell basis and write
 \begin{equation*}
\mathcal{B}^x = \{ \ket{b^x} = \bigotimes_{j=1}^{d} \ket{b^x_j} :   \ket{b^x_j} \in  \mathcal{B}_j\}.
\end{equation*}
The Hilbert space on which  the  QLGA evolves has as its   basis elements infinite tensor products  (over cell indices) of elements of $\mathcal{B}$ (basis of the cell Hilbert space).  These sequences, called the  {\it finite configurations},  consist of a finite region of cells in {\it active\/} states immersed in a background of cells in a fixed {\it quiescent\/} state.   For our purposes, we take  the quiescent state to be  $\ket{q} = \bigotimes^d \ket{0}$, while the rest are active. We write the  {\it finite configurations\/} basis of the QLGA Hilbert space, the {\it Hilbert space of finite configurations}, as
\begin{equation*} \mathcal{C} = \{ \bigotimes_{x \in \mathbb{Z}} \ket{b^{x}} :  \ket{b^{x}} \in \mathcal{B}^x,  \text{ all but finite number of the } \ket{b^{x}}  = \ket{q}\}.
\end{equation*}
Explicitly, each  $\ket{b^{x}} = \bigotimes_{j=1}^{d} \ket{b^x_{j}}$, where each    $\ket{b^x_{j}}\in  \mathcal{B}_j$ is a basis element of the subcell space $W_j$. $\mathcal{C}$ is thus  orthonormal under the inner product induced from $W_j$. This definition of  the   finite configurations  basis,  $\mathcal{C}$,  ensures that   it is  countable, so that the Hilbert space of finite configurations is separable  (in the topological sense). This Hilbert space  naming convention is adopted from~\cite{anw:odqca}; in the mathematics literature, it is called an {\it incomplete infinite tensor product\/} space, as in~\cite{ag:shrt,vn:idp}.

The QLGA evolves in two unitary steps:
 \begin{enumerate}[label=(\roman{*})] 
\item \label{propqlg} {\it Advection},  $\sigma$, that shifts the appropriate subcell value to the corresponding neighbor,
 \begin{equation}   \label{propeqn}
 \left . \begin{array} {cccc}
 \sigma : 
  \bigotimes_{x \in \mathbb{Z}}   \bigotimes_{j=1}^{d} \ket{b^x_{j}} \mapsto  \bigotimes_{x \in \mathbb{Z}}  \bigotimes_{j=1}^{d} \ket{b^{x+e_j}_{j}}.
    \end{array}
  \right .
\end{equation} 

\item \label{colqlg} {\it Scattering}, $\hat S$, which acts on each cell  by a local unitary scattering map $S: W \longrightarrow W$ that  fixes the quiescent state, i.e., $S(\ket{q}) = \ket{q}$,
\begin{equation*} 
\hat S:   \bigotimes_{x \in \mathbb{Z}}   \bigotimes_{j=1}^{d} \ket{b^x_{j}} \mapsto   \bigotimes_{x \in \mathbb{Z}}   S(\bigotimes_{j=1}^{d} \ket{b^x_{j}}).
\end{equation*}
 \end{enumerate}  
 Each time step,    the current state of  the QLGA is  mapped unitarily to the next by its {\it global evolution}, $\mathcal{G}$,
  \begin{equation*}  
\mathcal{G} = \sigma \hat S.
\end{equation*} 

This multiparticle model bears essentially the same relationship to a QRW as a second-quantized model does to a first-quantized single particle model:  it keeps track of the presence or absence of particles at the all the sites rather than the position of a single particle.  The underlying structure is an incomplete tensor product space and the evolution is defined by two operations.   One is advection (propagation) which is completely specified by cell decomposition into subcells and the corresponding neighborhood elements, and the other is scattering that happens at each site and  is the sole interaction amongst the particles.  Note that in this interpretation the active site model described in Subsection~\ref{subsec:qrwhistn} whose Hilbert space is given by Eq.~\eqref{eq:fqsq} is a hybrid between first and second quantized representations. The position of the particle has a first quantized representation, while the labels of the active sites have a second quantized representation. This already suggests that a fully second quantized picture, or QLGA, may have utility for QRW with history dependence, and it is this reformulation we turn to in the next section.

\section{QRW as QLGA}\label{sec:qrwqlga}
In general, a discrete-time QRW is a QLGA restricted to a single-particle subspace.  The simplest example of this is the original QLGA constructed by Meyer~\cite{bib:meyer2}.  We begin this section with a description of this model in the current context~\cite{bib:slwqcaqlga} and show how it restricts to a QRW. 

\begin{figure}[H] 
\includegraphics{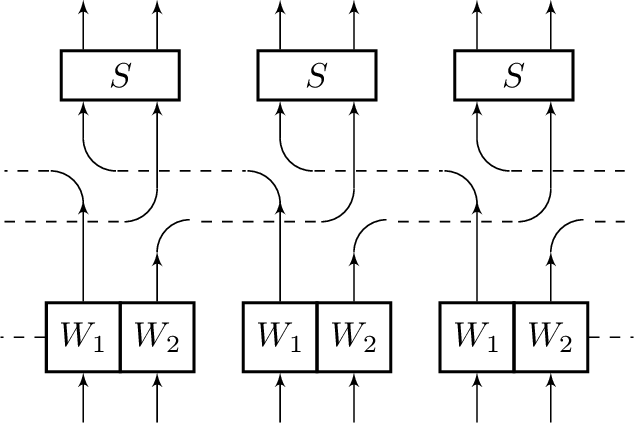}
\caption{QLGA model of the standard QRW, and of QRW with particle history.}
\label{figqq1}
\end{figure}

Let us recall the example QLGA of~\cite{bib:meyer2}.  The cell Hilbert space is $W =  W_1\otimes W_2 = \mathbb{C}^2\otimes\mathbb{C}^2$, with two subcells; each subcell space is a qubit, $W_1 = W_2 = \mathbb{C}^2$. The neighborhood is $\mathcal{E}=\{e_1 = +1, e_2= -1\}$ (the neighborhood elements $e_1$, $e_2$ seem opposite to the respective movement directions of the subcell elements, but this is required by  the expression for advection $\sigma$ in Eq.~\eqref{propeqn}). The local scattering operator $S$, in the  ordered basis of $W$,  $\{\vert00\rangle,\vert01\rangle,\vert10\rangle,\vert11\rangle\}$, is
\begin{equation}\label{eq:S1}
S=\begin{pmatrix}
1&0&0&0\\
0&ie^{i\alpha}\sin\theta&e^{i\alpha}\cos\theta&0\\
0&e^{i\alpha}\cos\theta&ie^{i\alpha}\sin\theta&0\\
0&0&0&e^{i\beta}\\
\end{pmatrix}.
\end{equation}
$S$ conserves particle number---since both $\ket{00}$ and $\ket{11}$ are eigenvectors, while the middle $2 \times 2$ block acts on the subspace  Span$\{ \ket{01}, \ket{10} \}$. This means that an evolution that begins in the single-particle sector will remain in the single particle sector, a fact that underpins the intrepretation of discrete time quantum random walks as a subclass of the dynamics of QLGA. The time evolution is illustrated in Fig.~\ref{figqq1}. 
 
Now let us interpret the state $\ket{01}$  at cell $x$ as the walker in a QRW at site $x$ moving to the right, i.e., in state $\ket{x}\ket{+1}$ at the end of the scattering step. In the same way, interpret the state $\ket{10}$ at cell $x$ as the walker in a QRW at site $x$ moving to the left, i.e., in state $\ket{x}  \ket{-1}$ at the end of the scattering step. 

Assume that the state of the QLGA is such that it is quiescent everywhere except at one cell $x$ where it is $\ket{01}$, i.e., it is in the single particle sector and is given by
\begin{equation*}
\ket{\psi}  =   \ldots   \ket{00}  \ket{00} \underbrace{\ket{01}}_{\text{$x$}}     \ket{00}  \ket{00} \ldots
\end{equation*}
Then after the advection step, it is
\begin{equation*}
\quad \sigma \ket{\psi}  =   \ldots  \ket{00}  \ket{00}   \ket{00} \underbrace{\ket{01}}_{\text{$x+1$}}     \ket{00}  \ket{00} \ldots
\end{equation*}
Similarly, if the initial state is
\begin{equation*}
\quad \ket{\psi} = \ldots\ket{00}\ket{00}\underbrace{\ket{10}}_{\text{$x$}}\ket{00}\ket{00}\ldots,
\end{equation*}
then after the advection step, it is
\begin{equation*}
\sigma \ket{\psi} = \ldots\ket{00}\underbrace{\ket{10}}_{\text{$x-1$}}\ket{00}\ket{00}\ket{00}\ldots
\end{equation*}
The scattering  operator, as mentioned, acts locally on each cell by $S$, and invariantly on the subspace Span$\{\ket{01},\ket{10}\}$, the $2 \times 2$ block representing the local scattering operator.  Hence the QRW is  captured completely by the single-particle sector of a QLGA, the space spanned by $\ket{\psi}$ restricted to the single particle sector, if the local scattering operator $S$ is as shown in Eq.~\eqref{eq:S1}.

The basic ingredients of the QRW as a restricted QLGA have been demonstrated by this simple example. Now that we have shown that a standard QRW model can be embedded in a QLGA, we proceed to the more complicated models with history. 

\subsection{QRW with particle history dependence as a QLGA} \label{subsec:qrwhistcqlg}

We treat the the Rohde, Brennen and Gilchrist model~\cite{rbg:qwwmrccf} described in Section~\ref{subsec:qrwhistc}~\ref{scmem}.  In this model, the particle retains a length $N$ history of velocities, and cyclically chooses the velocity to scatter.  The scattering matrix is a fixed symmetric matrix.  Brun, Carteret and Ambainis' models~\cite{bca:qwdmc,bca:qctrw} using $N$ different scattering matrices, one for each velocity (Section~\ref{subsec:qrwhistc}~\ref{scmem}), and Mc~Gettrick's model~\cite{mg:odqwwm}, with a record of $N$ previous sites (Section~\ref{subsec:qrwhistc}~\ref{snmem}), can be constructed similarly.  

We need to expand the Hilbert space of the subcells to include the history of velocities.  Note that this history is specific to the walker, and must hop with the walker. The cell Hilbert space thus must be $W = W_1\otimes W_2$, where $W_1 = W_2 = \mathbb{C}^2\otimes \bigotimes^{N}\mathbb{C}^2$. The first tensor factor of each subcell records the presence or absence of a left (respectively, right) moving walker, and the remaining $N$ additional factors are a record of the history of velocities, so the history of velocities hops with the walker.  Note that this keeps a velocity history of length $N+1$ instead of $N$ (including the current velocity encoded in the first factor).  The neighborhood is still $\mathcal{E}=\{e_1 = +1, e_2= -1\}$. Fig.~\ref{figqq1} applies to this case as well. 

The local scattering operator $S: W \longrightarrow W$ needs to account for the current dynamics as in the QRW without memory above, but must also do the book-keeping required to keep the history of the velocities. We can describe $S$ explicitly in terms of a local ricochet operator and a local memory operator.  Write a basis element of $W_1$ as $\ket{l} \ket{u_1 \ldots u_N}$, where $\ket{l}$ is the existence of a left-moving 
particle state, and $\ket{u_1 \ldots u_N}$ is the velocity history of that  left moving state.
 Each $\ket{u_i}$ is $\ket{+1}$ or $\ket{-1}$, with \ket{+1} representing a previous right move, and $\ket{-1}$ representing a previous left move. Similarly, for $W_2$, the analogous data are encoded in the basis element $\ket{r}\ket{v_1 \ldots v_N}$. A state of the cell can be written as
  \begin{equation}\label{eq:w1w2}
  \ket{l} \ket{u_1 \ldots u_N} \ket{r}\ket{v_1 \ldots v_N}   \in W_1\otimes W_2,
\end{equation}
which is isomorphic to
  \begin{equation*}
  \ket{l r} \ket{u_1 \ldots u_N}\ket{v_1 \ldots v_N}.
\end{equation*} 
We describe the scattering operation on the basis elements written in the form of Eq.~\eqref{eq:w1w2}. The ricochet operator  $R$ acts as
  \begin{equation*}  
 \left . \begin{array} {clll}
R : &W_1\otimes W_2 &\longrightarrow  & W_1\otimes W_2, \\
  &   \ket{0 0} \ket{u_1 \ldots u_N}\ket{v_1 \ldots v_N}   &\mapsto  &   \ket{0 0} \ket{u_1 \ldots u_N}\ket{v_1 \ldots v_N}, \\
  &   \ket{1 1} \ket{u_1 \ldots u_N}\ket{v_1 \ldots v_N}   &\mapsto  &   \ket{1 1} \ket{u_1 \ldots u_N}\ket{v_1 \ldots v_N}, \\
  & \ket{0 1} \ket{u_1 \ldots u_N}\ket{v_1 \ldots v_N}     &\mapsto  &\ket{0 1} \ket{u_1 \ldots u_N}\ket{v_1 \ldots v_{N-1}} U_b( \ket{v_N}),   \\
  & \ket{1 0} \ket{u_1 \ldots u_N}\ket{v_1 \ldots v_N}     &\mapsto  &\ket{1 0}  \ket{u_1 \ldots u_{N-1}} U_b( \ket{u_N})\ket{v_1 \ldots v_N},  \\
    \end{array}
  \right .
\end{equation*} 
where $U_b$ is a symmetric scattering matrix of the kind in Eq.~\eqref{velocityun}.  The memory operator $M$ acts as 
  \begin{equation*} 
   \left . \begin{array} {clll}
M : &W_1\otimes W_2 &\longrightarrow  & W_1\otimes W_2, \\
  &   \ket{0 0} \ket{u_1 \ldots u_N}\ket{v_1 \ldots v_N}   &\mapsto  &   \ket{0 0} \ket{u_1 \ldots u_N}\ket{v_1 \ldots v_N}, \\
  &   \ket{1 1} \ket{u_1 \ldots u_N}\ket{v_1 \ldots v_N}   &\mapsto  &   \ket{1 1} \ket{u_1 \ldots u_N}\ket{v_1 \ldots v_N}, \\
  &\ket{0 1} \ket{u_1 \ldots u_N}\ket{v_1 \ldots v_{N-1}, +1}     &\mapsto   &\ket{0 1} \ket{u_1 \ldots u_N}\ket{+1,v_1 \ldots v_{N-1}},   \\
  &\ket{0 1} \ket{u_1 \ldots u_N}\ket{v_1 \ldots v_{N-1}, -1}     &\mapsto   &\ket{1 0} \ket{+1,v_1 \ldots v_{N-1}}  \ket{u_1 \ldots u_N},  \\
  &\ket{1 0} \ket{u_1 \ldots u_{N-1}, +1}\ket{v_1 \ldots v_N}     &\mapsto     &\ket{0 1}\ket{v_1 \ldots v_N}   \ket{-1,u_1 \ldots u_{N-1}},  \\
  &\ket{1 0} \ket{u_1 \ldots u_{N-1}, -1}\ket{v_1 \ldots v_N}     &\mapsto     &\ket{1 0}  \ket{-1,u_1 \ldots u_{N-1}}  \ket{v_1 \ldots v_N}.
    \end{array}
  \right .
\end{equation*} 
Finally, the local scattering operator $S$ is
\begin{equation*} 
  S = M R.
\end{equation*} 
  
A basis state in the single particle sector of a QLGA corresponds to $\ket{l^x, r^x} = \ket{1 0}$  at exactly one cell $x$ (the superscript is the cell index), and $\ket{l^y, r^y} = \ket{0 0}$ for all $y \neq x \in \mathbb{Z}$.   Notice that this formulation suggests a generalization with a different scattering operator that still conserves the particle number (only allows states of this form), but, as a result of the scattering at the site, alters the internal state. For example, let us assume  that the post-scattering number state is   $\ket{l^x, r^x} = \ket{0 1}$, which means that  the left moving number state  $\ket{l^x}=\ket{0}$ is the ``particle absent" state. The scattering can be chosen so that the internal state $\ket{u^x_1 \ldots u^x_N}$  associated to  this  left moving  ``particle absent" state changes as it scatters, which then propagates as an ``external field'', creating vacuum-like modes still associated with the particle state but not hopping with the right moving $\ket{r^x}=\ket{1}$  ``particle present" state.
  
\subsection{QRW with site history dependence as a QLGA} \label{subsec:qrwhistcqlg}

In this subsection we show that it is also possible to capture models with site history dependence in a natural way as QLGA models.  Camilleri, Rohde and Twamley's QRW with site-history~\cite{crt:saqw} was described in Section~\ref{subsec:qrwhistn}.  This model maintains a history of visits to each site by the particle by keeping a qubit that records whether the site has been visited or not by acting on the qubit, each time a site is visited,  through a symmetric scattering matrix whose ``memory strength''  is the  parameter of the scattering matrix. Based on the record of a previous visit or not, the particle scatters differently through a controlled scattering matrix: scattering in a balanced manner without a prior visit, and scattering by a ``back-action'' scattering matrix if there has been a visit to the site.  For a finite lattice of size $N$, this requires a Hilbert space of dimension $N 2^{N+1}$ to capture the position, velocity and the record of visits to each node. Once again we note that this Hilbert space is a hybrid between first and second quantized representations.

The QLGA model encodes, in the cell Hilbert space, in addition to the velocity of the particle, the memory of whether each site has been visited in the past.  The cell Hilbert space therefore includes one more qubit than does the original QLGA we described at the beginning of this section~\cite{bib:meyer2,bib:slwqcaqlga}.  A cell has three subcells, and each subcell space is $W_0 = W_1 = W_2 = \mathbb{C}^2$, so the cell Hilbert space is $W = \mathbb{C}^2\otimes\mathbb{C}^2\otimes\mathbb{C}^2$.  Two of the subcells encode the walker's state as in the original QLGA, and the third subcell records the memory of visits to that site. Recall that the history of visits to a site belongs to the site itself; it does not hop with the walker. Thus, if we choose the  subcell $W_0$  to be the memory qubit, then we need to choose the neighborhood as $\mathcal{E}=\{e_0 = 0, e_1 = +1, e_2 = -1\}$, ensuring that subcell  $W_0$ does not shift. Advection effectively only acts on $W_1\otimes W_2$, shifting $W_1$ to the left and $W_2$ to the right. 

\begin{figure}[H] 
\includegraphics{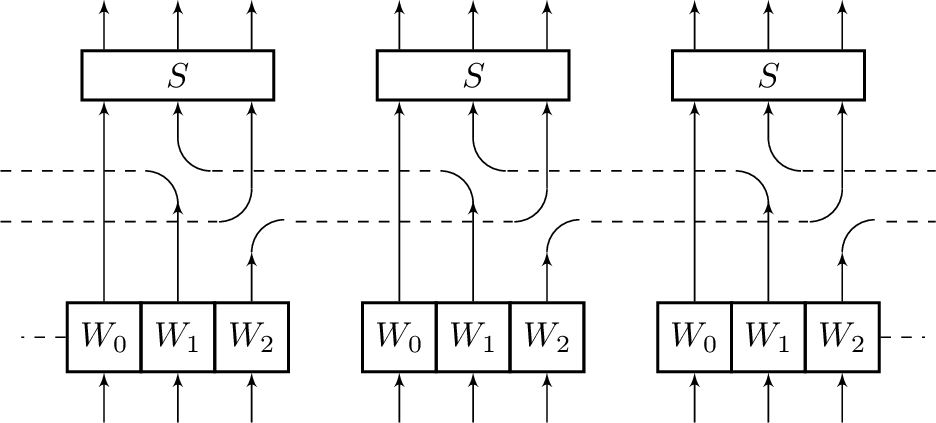}
\caption{A QLGA model of QRW with site history.}
\label{figqq2}
\end{figure}

The local scattering operator $S: W \longrightarrow W$ accounts for the current dynamics as before, and updates the memory subcell $W_0$ based on the state of $W_1\otimes W_2$. This is illustrated in Fig.~\ref{figqq2}.

As in the previous case, such a local scattering can be defined, and split into memory and ricochet operators.  Write a basis element of $W$ as $\ket{m}\ket{l}\ket{r}$, where $\ket{l} \in W_1$ is the presence of a left moving particle and $\ket{r} \in W_2$ is the presence of a right moving particle, and $\ket{m} \in W_0$ is the memory state, where $\ket{0}$ means ``not visited'' and $\ket{1}$ means ``have visited''.  Then the memory operator acts by
 \begin{equation*}  
 \left . \begin{array} {clll}
M : &W_0 \otimes W_1\otimes W_2 &\longrightarrow  & W_0 \otimes  W_1 \otimes W_2, \\
  &\ket{m}   \ket{0}    \ket{0}    &\mapsto  &\ket{m}   \ket{0}   \ket{0}, \\
  &\ket{m}   \ket{1}    \ket{1}    &\mapsto  &\ket{m}   \ket{1}   \ket{1}, \\
  &\ket{m}   \ket{0}    \ket{1}    &\mapsto  &U_M(\ket{m})   \ket{0}   \ket{1}, \\
  &\ket{m}   \ket{1}    \ket{0}    &\mapsto  &U_M(\ket{m})   \ket{1}   \ket{0},  \\
    \end{array}
  \right .
\end{equation*} 
where $U_M$ is as in Eq.~\eqref{memun}.

Now we describe  the action of local ricochet operator $R$ on the basis of $W$.
  \begin{equation*}  
 \left . \begin{array} {clll}
R : &W_0 \otimes W_1\otimes W_2 &\longrightarrow  & W_0 \otimes  W_1\otimes W_2, \\
  &\ket{0}   \ket{l}    \ket{r}    &\mapsto  &\ket{0}   R_0(\ket{l}   \ket{r}), \\
   &\ket{1}   \ket{l}    \ket{r}    &\mapsto  &\ket{1}   R_1(\ket{l}   \ket{r}), \\
    \end{array}
  \right .
\end{equation*} 
where $R_0$ is the local ``balanced'' ricochet operator (based on $U_0$ in Eq.~\eqref{fortun}), and in the  basis $\{\vert00\rangle,\vert01\rangle,\vert10\rangle,\vert11\rangle\}$ of $W_1\otimes W_2$ is
\begin{equation*}
R_0=\begin{pmatrix}
1&0&0&0\\
0&1/\sqrt{2}&i/\sqrt{2}&0\\
0&i/\sqrt{2} &1/\sqrt{2}&0\\
0&0&0&1\\
\end{pmatrix},
\end{equation*}
and the ``back action'' ricochet operator, $R_1$  (based on $U_1$ in Eq.~\eqref{bactun}), is
\begin{equation*}
R_1=\begin{pmatrix}
1&0&0&0\\
0&\cos\theta_b&i\sin\theta_b&0\\
0&i\sin\theta_b &\cos\theta_b&0\\
0&0&0&1\\
\end{pmatrix}.
\end{equation*}

\begin{figure}[H] 
\includegraphics{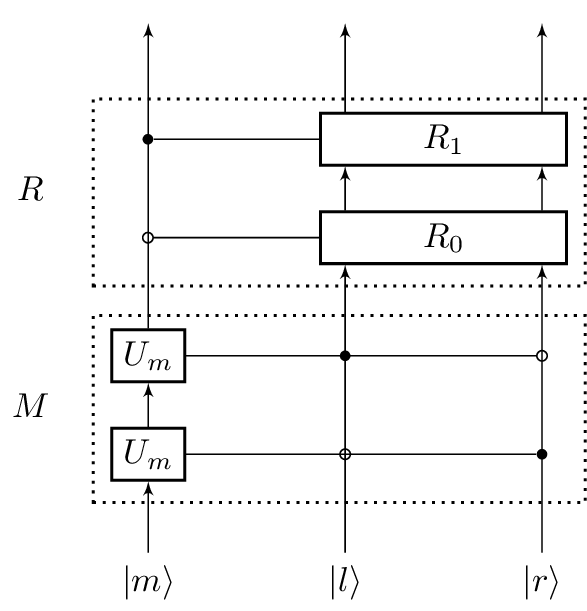}
\caption{Scattering operator $S$ as a sequence of controlled operations $M$ and $R$.}
\label{figqq3}
\end{figure}

Finally, the scattering operator $S$ is
  \begin{equation*} 
S =  R M.
  \end{equation*} 
Fig.~\ref{figqq3} shows $S$ as a sequence of controlled operations $M$ and $R$.
  
Thus the QLGA model elegantly and efficiently captures the structure and dynamics of QRW with site dependent history.  The Hilbert space has countable dimension even though the  lattice and hence the particle's position is unbounded. This contrasts with   the original QRW with site dependent history, in which the Hilbert space dimension grows exponentially with the size of the lattice. Moreover,  the QLGA model  provides the physical interpretation of a background quantum field interacting at every site with the particle. 

\section{Non-reversing and non-repeating QRW on a 2D lattice as QLGA}\label{sec:nrnr}

Proctor et al.\ construct a QRW on a two-dimensional lattice, with special class of scattering operators that have a memory in the sense that they either prevent the walker from going in the same direction for two consecutive steps ({\it non-repeating}), or ensure that the site just visited is not visited again at the next step ({\it non-reversing})~\cite{pbhmpbk:nrnrqw}.  In this section we show that this QRW\footnote{Notice that Rosmanis' one dimensional quantum ``snake'' model~\cite{bib:Rosmanis}, were a discrete time version constructed, could be encoded as a two dimensional QRW model with the two position components representing the location of the head and the tail, and the particle history tensor factor encoding the positions of the ``body'' of the ``snake''.  It would, of course, have different constraints than Proctor et al.'s model.} can also be realized as a QLGA model. 

\subsection{Non-reversing and non-repeating QRW on a two dimensional lattice}  \label{subsec:2dqrw}
The lattice is $\mathbb{Z} \times\mathbb{Z}$, so positions are denoted by $\ket{x}\ket{y} \in \ell^2(\mathbb{Z} \times \mathbb{Z})$, and velocities take values in a $4$-dimensional vector space  $V_p = \mathbb{C}^4$ in which the  basis elements $\ket{p} \in \{ \ket{w},\ket{e},\ket{s},\ket{n} \}$ signify the velocity directions west, east, south, north, respectively.  The Hilbert space for this case is
\begin{equation*}
\mathcal{H} = \ell^2(\mathbb{Z} \times \mathbb{Z}) \otimes \mathbb{C}^4 =  \ell^2(\mathbb{Z} \times \mathbb{Z}) \otimes V_p,
\end{equation*}
with typical basis elements written as $\ket{x}\ket{y}\ket{p}$.

A QRW evolves through two consecutive unitary actions on its state as in the original one-dimensional QRW.
 \begin{enumerate}[label=(\roman{*})] 
\item \label{proprr} Advection 
\begin{alignat*}{3}
A : 
&\ket{x}\ket{y}\ket{w} &&\mapsto &&\ket{x-1}\ket{y}\ket{w},  \\
&\ket{x}\ket{y}\ket{e} &&\mapsto &&\ket{x+1}\ket{y}\ket{e}, \\
&\ket{x}\ket{y}\ket{s} &&\mapsto &&\ket{x}\ket{y-1}\ket{s},\\
&\ket{x}\ket{y}\ket{n} &&\mapsto &&\ket{x}\ket{y+1}\ket{n}. \\
\end{alignat*}
\item \label{scatrr}  Scattering  
\begin{equation*}
S :     \ket{p} \mapsto    U_b(\ket{p}),
\end{equation*}
where $U_b$ is a unitary map on $\mathbb{C}^4$. When the diagonal of $U_b$ is all $0$s, this is called the {\it non-repeating\/} scattering operator  designated $C^{\neg{\rm rep}}$. It can also be taken to be of the form $C^{\neg{\rm rev}} = C^{\neg{\rm rep}} J$, where $J$,  in the ordered basis $\{\ket{w},\ket{e},\ket{s},\ket{n}\}$, is
\begin{equation*}
J=\begin{pmatrix}
0&1&0&0\\
1&0&0&0\\
0&0&0&1\\
0&0&1&0\\
\end{pmatrix}.
\end{equation*} 
$C^{\neg{\rm rev}}$ is called the {\it non-reversing\/} scattering operator.
\end{enumerate}   
The QRW transition $U$ is 
\begin{equation*}
U = A (I \otimes S).
\end{equation*}
This model of a QRW is notable besides being two-dimensional, as an example of a QRW that displays some memory simply by the choice of the scattering matrix.  To construct a QLGA for this model below, we introduce two-dimensional QLGA and reemphasize how a scattering matrix of a QRW embeds in a QLGA scattering rule. 

\subsection{QLGA model of the non-reversing and non-repeating QRW} \label{subsec:2dqrwqlg}
We generalize the one-dimensional QLGA model to a two dimensional lattice, and show how this captures the non-reversing and non-repeating QRW.  The lattice is $\mathbb{Z} \times\mathbb{Z}$. Since a particle can go in four directions, we need a neighborhood of four elements corresponding to the directions of the moves. Thus, the neighborhood is $\mathcal{E} = \{e_1=(1,0),  e_2=(-1,0),  e_3=(0,1), e_4=(0,-1)\}$, corresponding to the directions $\ket{w},\ket{e},\ket{s},\ket{n}$, respectively.  Each cell, instead of consisting of two copies of $\mathbb{C}^2$ recording the presence of a left or right moving particle, now has four copies of $\mathbb{C}^2$, comprising the cell Hilbert space $W = W_1\otimes W_2 \otimes W_3\otimes W_4$, with four subcells.  Each subcell space is $W_1 = W_2 = W_3 = W_4 = \mathbb{C}^2$ so that the basis elements $\ket{1000}, \ket{0100}, \ket{0010}, \ket{0001}$ represent the presence of a $\ket{w},\ket{e},\ket{s},\ket{n}$ moving particle.

The basis elements of the Hilbert space of the QLGA are thus
 \begin{equation*}  
  \bigotimes_{(x,y) \in \mathbb{Z} \times \mathbb{Z}}   \bigotimes_{j=1}^{4} \ket{b^{(x,y)}_j},
\end{equation*} 
where $\bigotimes_{j=1}^{4}  \ket{b^{(x,y)}_j}$ is the basis element of cell Hilbert space of the cell $(x,y) \in \mathbb{Z} \times\mathbb{Z}$, such that $ \ket{b^{(x,y)}_j} \in\{\ket{0},\ket{1}\} \subset W_j$. As defined above,  $W_j$ is the Hilbert space of the subcell $j$ of the cell $(x,y)$.

The evolution of  a QLGA, as in the one-dimensional lattice case, has two stages: 
 \begin{enumerate}[label=(\roman{*})] 
\item \label{prop2dqlg} {\it Advection},  $\sigma$, that shifts the appropriate subcell value to the corresponding neighbor,
 \begin{equation*}  
 \sigma : 
  \bigotimes_{(x,y) \in \mathbb{Z} \times \mathbb{Z}}   \bigotimes_{j=1}^{4}  \ket{b^{(x,y)}_j} \mapsto   \bigotimes_{(x,y) \in \mathbb{Z} \times \mathbb{Z}}   \bigotimes_{j=1}^{4}  \ket{b^{(x,y)+e_j}_j}.
\end{equation*}

\item \label{col2dqlg} {\it Scattering}, $\hat S$, which acts by a local unitary map $S: W \longrightarrow W$,  on each cell, 
\begin{equation*} 
\hat S:     \bigotimes_{(x,y) \in \mathbb{Z} \times \mathbb{Z}}   \bigotimes_{j=1}^{4}  \ket{b^{(x,y)}_j} \mapsto    \bigotimes_{(x,y) \in \mathbb{Z} \times \mathbb{Z}}   S \left(\bigotimes_{j=1}^{4}  \ket{b^{(x,y)}_j} \right).
\end{equation*}
The particular case we are considering requires that $S$ should mimic the action of $U_b$ in Section~\ref{subsec:2dqrw} above, so it is sufficient  to  take $S$  to be a block diagonal matrix in the standard basis of $W = \bigotimes^4\mathbb{C}^2$ such that it is exactly $U_b$ (either $C^{\neg{\rm rep}}$ or $C^{\neg{\rm rev}}$ as the case may be) acting on the block  spanned by the ordered basis elements 
$\{\ket{1000}, \ket{0100}, \ket{0010}, \ket{0001}\}$, and identity on the rest.

 \end{enumerate}  
 The QLGA global evolution $\mathcal{G}$ is
  \begin{equation*}  
\mathcal{G} =  \sigma \hat S.
\end{equation*} 
This example illustrates a two dimensional version of a QRW as a QLGA. We now see that at a local level the QLGA resource requirement, i.e., the cell Hilbert space dimension, increases exponentially with the dimension of the lattice, while globally, for any fixed lattice dimension, the QLGA Hilbert space's dimension (dimension of the Hilbert space of finite configurations) is still countable by construction.

\section{Conclusion}
In this paper, we examined several history dependent QRW and showed how they fit into a multiparticle model of QLGA.  This required us to classify the history dependent QRW into two categories:  QRW with dependence on particle history, and QRW with dependence on site history.  Doing so demonstrated that history dependent QRW are naturally interpretable as a multiparticle QLGA restricted to a single-particle sector interacting with a background of static particles.  The presence of the single particle, moving in a specific direction, is encoded in a product of qubits at each site, one for each allowed velocity.  The same number of additional tensor factors encode the memory state of the moving particle (or particles, since the model generalizes immediately to multiple moving particles with memory), while yet another tensor factor encodes the memory state of the site.  The latter tensor factors constitute what is essentially a background quantum field~\cite{bib:jlp}, with which the moving particle interacts.  This identification immediately suggested several generalizations to the history dependent QRW from which we started, and demonstrated that the site history models~\cite{crt:saqw} are quantum versions of well-studied classical models~\cite{bib:ruijgrokcohen}.
   
Moreover, the construction of QLGA we have used~\cite{bib:slwqcaqlga} also addresses one of the major concerns about the exponential growth of the dimension of Hilbert space in the model of QRW in~\cite{crt:saqw}.  We gave explicit constructions of QLGA equivalent to the main examples in each category of history dependent QRW.  This includes two dimensional models, e.g., the QLGA described in the last subsection is equivalent to QRW on a two dimensional lattice that are non-reversing (not revisiting the immediately previous site) or non-repeating (not repeating a hop in the same direction)~\cite{pbhmpbk:nrnrqw}. 

The physical interpretations of various history dependent QRW that become natural when the model is identified as a QLGA seem likely to provide at least heuristic, if not analytic, explanations for numerically observed behaviors like ``anomalous'' quantum diffusion and decoherence/transition to classicality.  These improved understandings may motivate construction of novel quantum algorithms, possibly analogous to classical algorithms based upon history dependent random walks.  Furthermore, since QLGA models are not restricted to the single (moving) particle sector, they provide a natural framework within which to extend history dependent QRW to multiple interacting walkers.  History {\it independent\/} QRW with multiple particles are capable of universal quantum computation~\cite{bib:cgz}, so there are likely to be quantum algorithmic applications for multiparticle sectors of QLGA encoding history dependent walkers.

\def\MR#1{\relax\ifhmode\unskip\spacefactor3000 \space\fi
  \href{http://www.ams.org/mathscinet-getitem?mr=#1}{MR#1}}


\begin{bibdiv}
\begin{biblist}
 
\bib{bib:pearson}{article}{
      author={Pearson, K.},
       title={The problem of the random walk},
        date={1905},
     journal={Nature},
      volume={72},
       pages={294},
}

\bib{bib:raleigh1880}{article}{
      author={Raleigh},
       title={On the resultant of a large number of vibrations of the same pitch and of arbitrary phase},
        date={1880},
     journal={Philosophical Magazine},
      volume={10},
       pages={73\ndash78},
}

\bib{bib:raleigh1905}{article}{
      author={Raleigh},
       title={The problem of the random walk},
        date={1905},
     journal={Nature},
      volume={72},
       pages={318},
}

\bib{bib:bachelier}{article}{
      author={Bachelier, L.},
       title={{\it Th\'eorie de la sp\'eculation}},
        date={1900},
     journal={{\it Annales Scientifiques de l'{\'E}cole Normale Sup\'erieure}},
      volume={3},
       pages={21\ndash86},
}

\bib{bib:polya1}{article}{
      author={P\'olya, G.},
       title={{\it Wahrscheinlichkeitstheoretisches \"uber die `Irrfahrt'}},
        date={1919},
     journal={{\it Mitteilungen der Physikalischen Gesellschaft Z\"urich}},
      volume={19},
       pages={75\ndash86},
}

\bib{bib:polya2}{article}{
      author={P\'olya, G.},
       title={{\it Quelques probl\`emes de probabilit\'e se rapportant \`a la `promenade au hasard'}},
        date={1919},
     journal={{\it l'Enseignement Math\'ematique}},
      volume={20},
       pages={444\ndash445},
}

\bib{bib:metropolisulam}{article}{
       title={The Monte Carlo method},
      author={Metropolis, N.},
      author={Ulam, S.},
     journal={Journal of the American Statistical Association},
      volume={44},
       pages={335\ndash341},
        year={1949},
}

\bib{bib:papadimitriou}{inproceedings}{
      author={Papadimitriou, C. H.},
       title={On selecting a satisfying truth assignment},
   booktitle={Proceedings of the 32nd Annual Symposium on Foundations of Computer Science},
        date={1991},
       pages={163\ndash169},
}

\bib{bib:schoning}{inproceedings}{
      author={Sch\"oning, U.},
       title={A probabilistic algorithm for $k$-SAT and constraint satisfaction problems},
   booktitle={Proceedings of the 40th Annual Symposium on Foundations of Computer Science},
        year={1999},
       pages={410\ndash414},
}

\bib{bib:meyer2}{article}{
      author={Meyer, D. A.},
       title={From quantum cellular automata to quantum lattice gases},
        date={1996},
     journal={Journal of Statistical Physics},
      volume={85},
       pages={551\ndash574},
}

\bib{bib:abnvw}{inproceedings}{
      author={Ambainis, A.},
      author={Bach, E.},
      author={Nayak, A.},
      author={Vishwanath, A.},
      author={Watrous, J.},
       title={One-dimensional quantum walks},
   booktitle={Proceedings of the 33rd Annual ACM Symposium on the Theory of Computing},
      series={STOC '01},
        year={2001},
    location={Hersonissos, Greece},
       pages={37\ndash49},
   publisher={ACM},
     address={New York, NY},
} 

\bib{bib:boghosian1998}{article}{
	author = {Boghosian, B. M.},
	author = {Taylor, W.},
	title = {Simulating quantum mechanics on a quantum computer},
	journal = {Physica D--Nonlinear Phenomena},
	year = {1998},
	volume = {120},
	%number = {1\ndash2},
	pages = {30\ndash42},
}

\bib{bib:grover}{inproceedings}{
      author={Grover, L.},
       title={A fast quantum mechanical algorithm for database search},
   booktitle={Proceedings of the 28th Annual ACM Symposium on the Theory of Computing},
        year={1996},
    location={Philadelphia, PA},
       pages={212\ndash219},
   publisher={ACM},
     address={New York, NY},
} 

\bib{bib:skw}{article}{
      author={Shenvi, N.},
      author={Kempe, J.},
      author={Whaley, K. B.},
       title={Quantum random walk search algorithm},
     journal={Physical Review A},
      volume={67},
        year={2003},
       pages={052307/1\ndash11}
}

\bib{bib:akrcmqwf}{inproceedings}{
      author={Ambainis, A.},
      author={Kempe, J.},
      author={Rivosh, A.},
       title={Coins make quantum walks faster},
   booktitle={Proceedings of the 16th ACM-SIAM Symposium on Discrete Algorithms},
   %   volume={70},
        year={2005},
       pages={1099\ndash 1108}
}

\bib{bib:farhigutmann}{article}{
      author={Farhi, E.},
      author={Gutmann, S.},
       title={Analog analogue of a digital quantum computation},
     journal={Physical Review A},
      volume={57},
        year={1998},
       pages={2403\ndash2406}
}

\bib{bib:childsgoldstone1}{article}{
      author={Childs, A. M.},
      author={Goldstone, J.},
       title={Spatial search by quantum walk},
     journal={Physical Review A},
      volume={70},
        year={2004},
       pages={022314/1\ndash11}
}

\bib{bib:childsgoldstone2}{article}{
      author={Childs, A. M.},
      author={Goldstone, J.},
       title={Spatial search and the Dirac equation},
     journal={Physical Review A},
      volume={70},
        year={2004},
       pages={042312/1\ndash5}
}

\bib{bib:farhigoldstonegutmann}{article}{
      author={Farhi, E.},
      author={Goldstone, J.},
      author={Gutmann, S.},
       title={A quantum algorithm for the Hamiltonian NAND tree},
     journal={Theory of Computing},
      volume={4},
        year={2008},
       pages={169\ndash190}
}

\bib{bib:childesclevedeotto}{inproceedings}{
      author={Childs, A. M.},
      author={Cleve, R.},
      author={Deotto, E.},
      author={Farhi, E.},
      author={Gutmann, S.},
      author={Childs, A. M.},
       title={Exponential algorithmic speedup by a quantum walk},
   booktitle={Proceedings of the $35$th ACM Symposium on the Theory of Computing},
      %volume={4},
        year={2003},
       pages={59\ndash68},
   publisher={ACM},
     address={New York, NY},
}

\bib{bib:orr}{article}{
      author={Orr, W. J. C.},
       title={Statistical treatment of polymer solutions at infinite dilution},
     journal={Transactions of the Faraday Society},
      volume={43},
        year={1947},
       pages={12\ndash27}
}

\bib{bib:flory}{article}{
      author={Flory, P. J.},
       title={The configuration of real polymer chains},
     journal={The Journal of Chemical Physics},
      volume={17},
        year={1949},
       pages={303\ndash310}
}

\bib{bib:kuhn}{article}{
      author={Kuhn, W.},
       title={{\it \"Uber die Gestalt fadenf\"ormiger Molek\"ule in L\"osungen}},
     journal={{\it Kolloid-Zeitschrift}},
      volume={68},
        year={1934},
       pages={2\ndash15}
}

\bib{bib:floryfox}{article}{
      author={Flory, P. J.},
      author={Fox Jr., T. G.},
       title={Treatment of intrinsic viscosities},
     journal={Journal of the American Chemical Society},
      volume={73},
        year={1951},
       pages={1904\ndash1908}
}

\bib{bib:lawler}{article}{
      author={Lawler, G. F.},
       title={A self-avoiding random walk},
     journal={Duke Mathematical Journal},
      volume={47},
        year={1980},
       pages={655\ndash693}
}

\bib{bib:pemantle}{article}{
      author={Pemantle, R.},
       title={Choosing a spanning tree for the integer lattice uniformly},
     journal={The Annals of Probability},
      volume={19},
        year={1991},
       pages={1559\ndash1574}
}

\bib{bib:wilson}{inproceedings}{
      author={Wilson, D. B.},
       title={Generating random spanning trees more quickly than the cover time},
   booktitle={Proceedings of the 28th Annual ACM Symposium on the Theory of Computing},
        year={1996},
    location={Philadelphia, PA},
       pages={296\ndash303},
   publisher={ACM},
     address={New York, NY},
} 

\bib{bib:proppwilson}{article}{  
      author={Propp, J. G.},
      author={Wilson, D. B.},
       title={How to get a perfectly random sample from a generic
              Markov chain and generate a random spanning tree of
              a directed graph},
     journal={Journal of Algorithms},
        year={1998},
      volume={27},
       pages={170\ndash217},
}

\bib{bca:qwdmc}{article}{
       title={Quantum walks driven by many velocities},
      author={Brun, T. A.},
      author={Carteret, H. A.},
      author={Ambainis, A.},
     journal={Physical Review A},
      volume={67},
       issue={5},
       pages={052317/1\ndash17},
        year={2003},
%        note={\href{http://dx.doi.org/10.1103/PhysRevA.67.052317}{doi:10.1103/PhysRevA.67.052317}},
}

\bib{bca:qctrw}{article}{
       title={Quantum to classical transition for random walks},
      author={Brun, T. A.},
      author={Carteret, H. A.},
      author={Ambainis, A.},
     journal={Physical Review Letters},
      volume={91},
       issue={13},
       pages={130602/1\ndash4},
        year={2003},
}

\bib{mg:odqwwm}{article}{
      author={Mc~Gettrick, M.},
       title={One dimensional quantum walks with memory},
        date={2010},
     journal={Quantum Information and Computation},
      volume={10},
       issue={5\&6},
       pages={0509\ndash0524},
}

\bib{rbg:qwwmrccf}{article}{
       title={Quantum walks with memory provided by recycled velocities and a memory of the velocity-flip history},
      author={Rohde, P. P.},
      author={Brennen, G. K.},
      author={Gilchrist, A.},
     journal={Physical Review A},
      volume={87},
       issue={5},
       pages={052302/1\ndash11},
        year={2013},
%        note={\href{http://dx.doi.org/10.1103/PhysRevA.87.052302}{doi:10.1103/PhysRevA.87.052302}}
}

\bib{pbhmpbk:nrnrqw}{article}{
       title={Non-reversal and non-repeating quantum walks},
      author={Proctor, T.},
      author={Barr, K.},
      author={Hanson, B.},
      author={Martiel, S.},
      author={Pavlovic, V.},
      author={Bullivant, A.},
      author={Kendon, V.},
        year={2013},
     journal={preprint},
        note={\href{http://arxiv.org/abs/1303.1966}{arXiv:1303.1966}},
}

\bib{crt:saqw}{article}{
      author={Camilleri, C.},
      author={Rohde, P. P.},
      author={Twamley, J.},
       title={Self-avoiding quantum walks},
        date={2014},
     journal={preprint},
        note={\href{http://arxiv.org/abs/1401.1869}{arXiv:1401.1869}},
}

\bib{bib:Rosmanis}{article}{
  title = {Quantum snake walk on graphs},
  author = {Rosmanis, A.},
  journal = {Physical Review A},
  volume = {83},
  issue = {2},
  pages = {022304},
  numpages = {14},
  year = {2011},
}

\bib{bib:slwqcaqlga}{article}{
      author={Shakeel, A.},
      author={Love, P.},
       title={When is a Quantum Cellular Automaton (QCA) a Quantum Lattice Gas Automaton (QLGA)?},
        date={2013},
     journal={Journal of Mathematical Physics},
      volume={54},
       pages={092203/1\ndash40},
}

\bib{bib:schrodinger1}{article}{
       title={Uber die kr\"aftefreie Bewegung in der relativistischen Quantenmechanik},
      author={Schr\"odinger, E.},
        date={1930}, 
     journal={Sitzungsberichte der Preu{\ss}ischen Akademie der Wissenschaften.  Physikalisch-mathematische Klasse},
       pages={418\ndash428},
}

\bib{bib:schrodinger2}{article}{
       title={{Zur Quantendynamik des Elektrons}},
      author={Schr\"odinger, E.},
        date={1931}, 
     journal={Sitzungsberichte der Preu{\ss}ischen Akademie der Wissenschaften. Physikalisch-mathematische Klasse},
       pages={63\ndash72},
}

\bib{bib:jlp}{article}{
      author={Jordan, S. P.},
      author={Lee, K. S. M.},
      author={Preskill, J.}
       title={Quantum algorithms for quantum field theories},
        date={2012},
     journal={Science},
      volume={336},
       pages={1130\ndash1133},
}

\bib{bib:ruijgrokcohen}{article}{
      author={Ruijgrok, Th.~W.},
      author={Cohen, E. D. G.},
       title={Deterministic lattice gas models},
        date={1988},
     journal={Physics Letters A},
      volume={133},
       pages={415\ndash 418},
}

\bib{anw:odqca}{inproceedings}{
      author={Arrighi, P.},
      author={Nesme, N.},
      author={Werner, R.},
       title={One-dimensional quantum cellular automata over finite, unbounded
              configurations},
        year={2008},
   booktitle={Language and Automata Theory and Applications},
 seriestitle={Lecture Notes in Computer Science},
      volume={5196},
       pages={64\ndash75},
   publisher={Springer},
     address={Berlin},
} 

\bib{vn:idp}{article}{
      author={von Neumann, J.},
       title={On infinite direct products},
        date={1939},
     journal={Composito Mathematica},
      volume={6},
       pages={1\ndash 77},
}

\bib{ag:shrt}{book}{
      author={Guichardet, A.},
       title={Symmetric Hilbert Spaces and Related Topics},
        date={1972},
      series={Lecture Notes in Mathematics},
      volume={261},
   publisher={Springer},
     address={Berlin},
}

\bib{bib:cgz}{article}{
      author={Childs, A. M.},
      author={Gosset, D.},
      author={Webb, Z.}
       title={Universal computation by multiparticle quantum walk},
        date={2013},
     journal={Science},
      volume={339},
       pages={791\ndash794},
}





\end{biblist}
\end{bibdiv}
\end{document}